# A new perspective on materials for plasmonics

Babak Dastmalchi[1,2*], Philippe Tassin[3], Thomas Koschny[1],

and Costas M. Soukoulis[1,2]

[1] Ames Laboratory—U.S. DOE and Department of Physics and Astronomy, Iowa State University, Ames, IA 50011, USA

[2] Institute of Electronic Structure and Lasers (IESL), FORTH, 71110 Heraklion, Crete, Greece

[3] Department of Applied Physics, Chalmers University, SE-412 96 Göteborg, Sweden

**Surface plasmon polaritons are electromagnetic waves propagating on the surface of a metal. Thanks to subwavelength confinement to the surface, they can concentrate optical energy on the micrometer or even nanometer scale, enabling new applications in bio-sensing, optical interconnects, and nonlinear optics, where small footprint and strong field concentration is of the essence. The major obstacle in developing plasmonic applications is dissipative loss, which limits the propagation length of surface plasmons and broadens the bandwidth of surface-plasmon resonances. Here we present a new analysis of plasmonic materials and geometries that fully considers the trade-off between the propagation length and the degree of confinement of surface plasmon polaritons and allows a fair comparison between different substrates. It is based on a two-dimensional analysis of two independent figures-of-merit and we apply the analysis to all relevant plasmonic materials, e.g., noble metals, aluminium, silicon carbide, doped semiconductors, graphene, etc. Our analysis substantially eases the selection of a plasmonic material for a specific application, and provides guidance on what parameters should be tuned to improve performance of a particular application.**

The propagation of electromagnetic waves at the surface of conducting media was studied as early as the beginning of the 20[th] century by Zeneck[1] and Sommerfeld[2] in the framework of the interaction of radio waves with the Earth. However, it was the work of Otto[3,4], Economou[5] and Kretschmann[6] that started the modern plasmonics field by providing a detailed theoretical description and experimental methods to excite surface plasmon polaritons on films of noble metals. Because of advances in nanofabrication, characterization and computational electromagnetism, plasmonics has moved to the forefront of photonics during the last decades [7–15]. Indeed, the ability to squeeze optical energy into subwavelength volumes[16] makes surface plasmon modes attractive candidates for use in applications like bio-sensing[17,18], nonlinear optics[19,20] and solar energy[21], where the extreme localization of electromagnetic fields can improve efficiency, and in optical interconnects and chips[22–25], where the strong confinement results in miniaturization and dense packing of electrical as well as optical elements with small footprint.

Nevertheless, the development of applications based on plasmonics is hindered by dissipative loss. Because of conversion of optical energy to heat[26], surface plasmons decay while propagating along the surface. The propagation length usually depends not only on the material properties of the supporting medium, but also on its geometry, the frequency of operation, and the field symmetry of the plasmon mode. However, adjusting these degrees of freedom as, for example, in long-range surface plasmon modes[27] or using V-grooves[28], normally results in a trade-off between propagation length and level of confinement. Therefore, researchers have recently started exploring new low-loss plasmonic materials[29–34], including doped semiconductors, metal alloys, and graphene[35–39]. The aim of this Article is to develop a coherent method to compare plasmonic materials, carefully taking the trade-off between the propagation length and the degree of confinement into consideration, and to provide a comprehensive overview of the performance of current bleeding edge plasmonic materials using this universal framework.

**RESULTS**

Our method is based on a two-dimensional relation of the two fundamental and competing aspects of all surface plasmon waves: propagation length and degree of confinement. Let us, therefore, first define how we will measure these two parameters. Various definitions for the degree of confinement in a plasmonic waveguide are encountered in the literature, either based on the spatial extent of the energy[40,41] or on the distance over which the field decays from its local maximum in a guiding geometry[42]. While they differ in how they treat the near-field structure, they have in common that the confinement is predominantly determined by the exponential decay of the fields outside the plasmonic waveguide. The most intuitive definition for a figure-of-merit (FOM) for the degree of confinement is, then, the inverse of the distance over which the fields drop to 1/e of their maximum amplitude when moving away from the supporting interface, normalized by the free-space wavelength $\lambda_0$. If $k_\perp$ is the modal wave number in the transverse direction, then the figure-of-merit is given by

$$\delta = \frac{1}{\text{Im}[k_\perp]}$$

$$\text{FOM}_{\text{conf}} = \frac{\lambda_0}{\delta} \qquad (1)$$

This figure-of-merit measures the confinement of the surface wave with respect to the corresponding vacuum wavelength. There is more conformity in the literature about how to measure the propagation distance of a surface-plasmon polariton $L_{\text{prop}}$. It is usually defined as the distance over which the mode can propagate along the supporting interface until the field amplitudes drop to 1/e of its initial magnitude, now normalized by the surface-plasmon polariton wavelength $\lambda_{\text{spp}}$:

$$L_{\text{prop}} = \frac{1}{\text{Im}[k_\parallel]}$$
$$\text{FOM}_{\text{prop}} = \frac{L_p}{\lambda_{\text{spp}}} = \frac{\text{Re}[k_\parallel]}{2\pi\,\text{Im}[k_\parallel]} \qquad (2)$$

This figure-of-merit can effectively be thought of as how many wavelengths a surface-plasmon polariton mode can travel until its field amplitude has died out. Evidently, both figures-of-merit depend on the dispersion relation of the plasmon mode and, hence, they are influenced by material loss, the specific geometry of the waveguide, and the field symmetry of the plasmon mode. It is desirable to have both FOMs as large as possible for general plasmonic application. Unfortunately, this can usually not be achieved simultaneously.

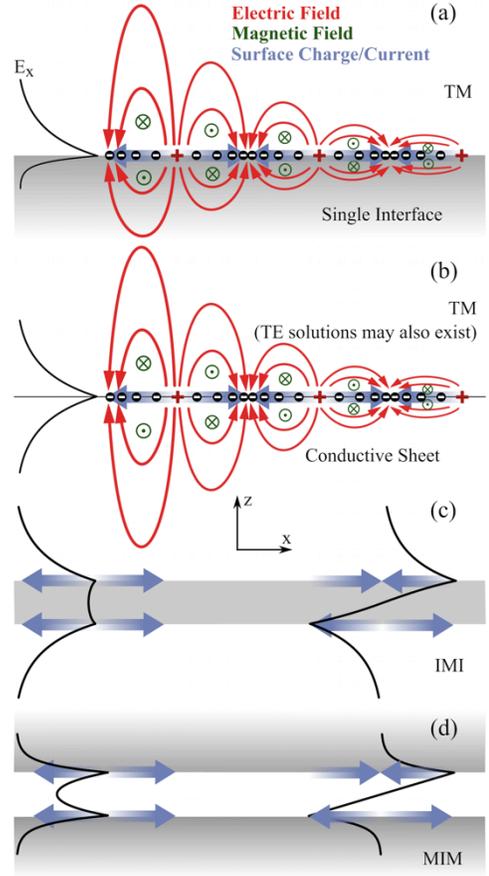

**Figure 1 | Schematic geometries and field/current distributions. a**, Single interface, **b**, Conductive sheet, **c**, Insulator-Metal-Insulator (IMI). **d**, Metal-Insulator-Metal.

### DRUDE MATERIALS

Surface plasmon polaritons are essentially transverse magnetic (TM) bound modes propagating at the interface of materials with permittivities of opposite sign (real part)—see the illustration in figure 1a for a schematic of the fields and the currents. The majority of materials with a negative permittivity—including the popular noble metals, transparent conducting oxides, and doped

semiconductors—can be fairly accurately modelled by a simple free-electron model, known as the Drude model, valid in frequency bands below the interband transition frequencies. Therefore, it will be insightful to first understand the general plasmonic behaviour of a material with Drude response:

$$\varepsilon(\omega) = 1 - \frac{\omega_p^2}{\omega(\omega+i\Gamma)} \quad (3)$$

where $\omega_p^2 = n e^2 / (\varepsilon_0 m^*)$ and $\Gamma$ are the plasma and collision frequencies, $n$ is the density of the free charge carriers in the material, and $e$ and $m^*$ are the electrical charge and effective mass of the carriers, respectively. Figure 2a shows a logarithmic plot of the frequency-dependent permittivity of a Drude material and the dispersion relation of the surface-plasmon polaritons on its surface. We can recognize four distinct generic regions: region I below the collision frequency, where the metal is primarily a dissipative conductor, region II below the plasma frequency, where the metal is dominated by polarization current and becomes primarily as a lossy negative dielectric, region III around the plasma frequency, and region IV above the plasma frequency, where the metal is a positive lossy dielectric with superluminal phase velocity. In figure 2b, we show the two-dimensional FOM graph with the propagation figure-of-merit on the horizontal axis and the confinement figure-of-merit on the vertical axis. Note that the low-frequency and high-frequency regions can be approximated by simple power laws, shown by thin lines in figure 2b, from which the trade-off between confinement and propagation length is evident. At the lowest frequencies, surface plasmon waves have a very high propagation figure-of-merit, but since they are closely following the light line, they are only weakly confined to the interface—these solutions are commonly known as Zenneck-Sommerfeld waves[1,2]. The most interesting region for plasmonics is undoubtedly region IIIa, i.e., frequencies close to but still below $\omega_{spp} = \omega_p/\sqrt{2}$. Here the plasmonic modes exhibit the strongest subwavelength nature and the best values for the figures-of-merit are found in this region. As the frequency approaches $\omega_{spp}$, the surface modes become very lossy, reducing the propagation distance to fractions of an SPP wavelength especially for the part with negative slope (shown by the dashed line in the FOM graph of figure 2b). Once the dispersion curve crosses the light-line, modes become leaky and energy is radiated away from the surface. Finally, in the high-frequency limit, the propagation distance recovers, nevertheless, due to the leaky nature, solutions are no longer confined to the interface. The plasmons in region IIIa always offer the best confinement-propagation trade-off.

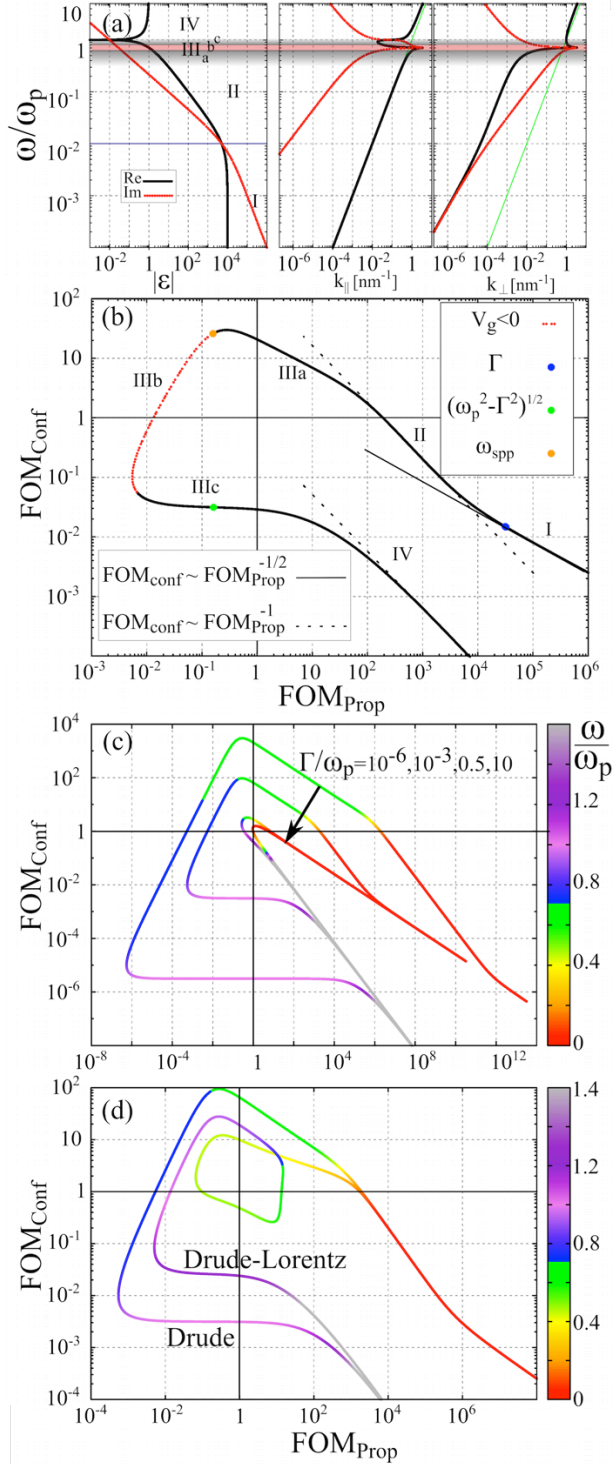

**Figure 2 | Dispersion relations and figures-of-merit. a,** From left to right: real (black) and imaginary (red) parts of the permittivity, the parallel component of the wave vector, and the normal component of the wave vector. The green lines in the dispersion diagrams mark the vacuum light-line. **b,** The behavior of the two FOMs for a conductor with Drude response. **c,** The dependence of the FOMs on the ratio of the collision frequency $\Gamma$ to the plasma frequency $\omega_p$. **d,** Comparison of FOM relations for a pure Drude, and a Drude-Lorentz dielectric function. Drude part for both curves is similar. The Lorentz term is centered at $\omega_0 = 0.6\ \omega_p$ with a width of $\Gamma = 1\times10^{-2}\ \omega_p$, an oscillator strength of $\omega_L = 0.7\ \omega_p$, and causes the additional loop in the graph.

For a Drude material, there is essentially only one parameter changing the shape of the two-dimensional FOM graph: the ratio of the collision frequency over the plasma frequency. Figure 2c demonstrates the impact of variations of this ratio on the FOM graph. By lowering the collision frequency (essentially reducing the average momentum lost in collisions), the area enclosed by the FOM line expands to achieve a better trade-off between confinement and propagation distance. It will, therefore, be desirable to choose materials with an as small as possible $\Gamma/\omega_p$ ratio. Choosing a material with a different plasma frequency, while keeping $\Gamma/\omega_p$ constant, allows shifting the frequency of operation.

The plasma frequency being proportional to the square root of the carrier density is also of importance for tunability. Bulk metals do not easily permit changes in the carrier concentration. However, for some other classes of materials, such as nonstoichiometric compounds,[43] photon-induced conductors[44] and graphene, the carrier concentration can readily be changed by several orders of magnitude with electrical[45,46], optical, or other methods of doping in order to tailor the plasmonic properties. Increasing the carrier concentration will generally have two effects: (1) It will expand the FOM graph to better figures-of-merit in region IIIa, and (2) the frequency band of operation will be blueshifted.

At visible frequencies, interband transitions play an important role in most conductors, and a simple Drude model can no longer represent their optical response. The effect of interband transitions can be modelled by adding multiple resonators represented by Lorentz terms[47]. In figure 2d, we show the FOM graph for a conductor with a single interband transition frequency. The FOM graph makes a loop with detrimental effect on the FOM-graphs. It is generally better to stay clear of interband transitions in plasmonic materials.

**SUMMARY OF REAL MATERIALS**

Having understood the properties of surface plasmon polaritons at the surface of a Drude material, we can now plot the two-dimensional FOM graph of experimental data for state of the art plasmonic materials. Figure 3 shows the FOM graph for experimental data of select bulk materials. The permittivity of noble metals (Ag, Au, and Cu) are obtained from reference 48. Data for Al and Be are from reference 49. Permittivity of transparent conductive oxides AZO, GZO, ITO are extracted from figure 10, and permittivity of ZrN and TiN are extracted from figure 15 of reference 50, respectively. Multiple data sets may exist for each material with varying deposition or measurement conditions. Additional materials are provided in the supplementary material[51].

To understand the behaviour of the materials in figure 3, the previous discussion of generic Drude materials can be readily used as a guide. For example, Be and Al have the highest plasma frequencies of all plasmonic materials and, therefore, region III of their FOM curves appear at very small free-space wavelengths in the ultraviolet range[i] (note the color bar indicating the wavelength). On the contrary, the FOM curves for Cu, Au, and Ag start from considerably longer wavelengths due to their smaller plasma frequencies. Similarly, the reason for the good performance of TCOs in the infrared region of the electromagnetic spectrum lies in their very low plasma frequency.

---

[i] As found from the fitted Drude functions for these real metals—see, for instance, table I from references 47 and 59).

Likewise, for the metal nitrides, TiN performs better in near infrared (lower plasma frequency), while ZrN accommodates the visible range (higher plasma frequency).

We now focus on the collision frequency. For the noble metals, the same trend as in figure 2b can be observed for the experimental material data. Ag, with the lowest collision frequency, turns out to be the best performing material, and Au and Cu are performing only slightly worse. For frequencies in the far infrared, yet another group of materials[49] attaining negative permittivity becomes important. Here, a bound electronic response (Lorentz type) rather than a free electronic response (Drude type) occurs due to a phonon-polariton interaction (see figure 4a). In this frequency range, silicon carbide shows excellent performance, effectively outperforming the response of silver in the visible.

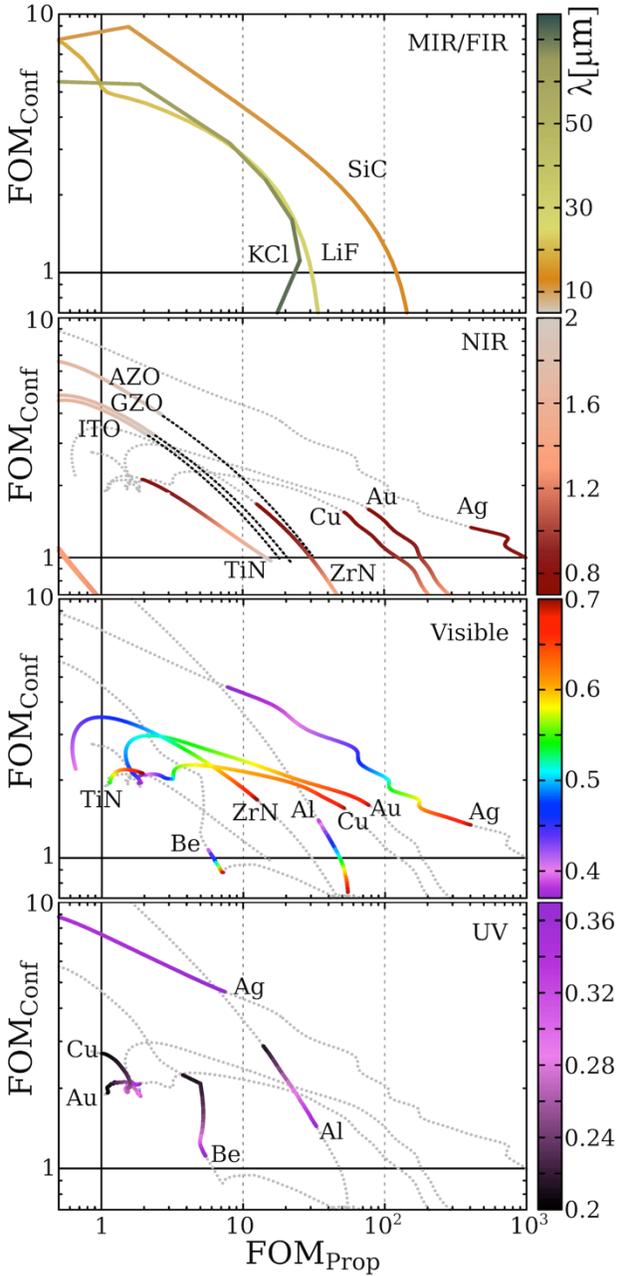

**Figure 3 | Figure-of-merit diagrams of single interface surface plasmon polaritons for experimental data in different regions of electromagnetic spectrum.** Plots are color-coded based on wavelength (μm). Gray dotted lines show the continuation of the dispersion out of the designated wavelength range. Black dashed lines in IR range are extrapolation of TCOs from 2μm to 5μm.

## TWO-DIMENTIONAL CONDUCTING SHEETS AND GRAPHENE

The properties of surface plasmons can be altered by changing the geometry, e.g. having them propagate on a thin film of a conducting material rather than at the interface of a bulk material. Thin films with finite thickness will later be explored in the next section. Here, we first look at a special limit of a thin film, namely a two-dimensional sheet of conductive material (see figure 1b for a schematic of the plasmon polariton mode). One example of such a two-dimensional plasmonic system is graphene, which consists of a single layer of covalently bonded carbon atoms. Another example is a conductive surface created by photon-excited carriers in a semiconductor like GaAs. A two-dimensional supporting medium for SPPs can be best modelled by a current sheet[51].

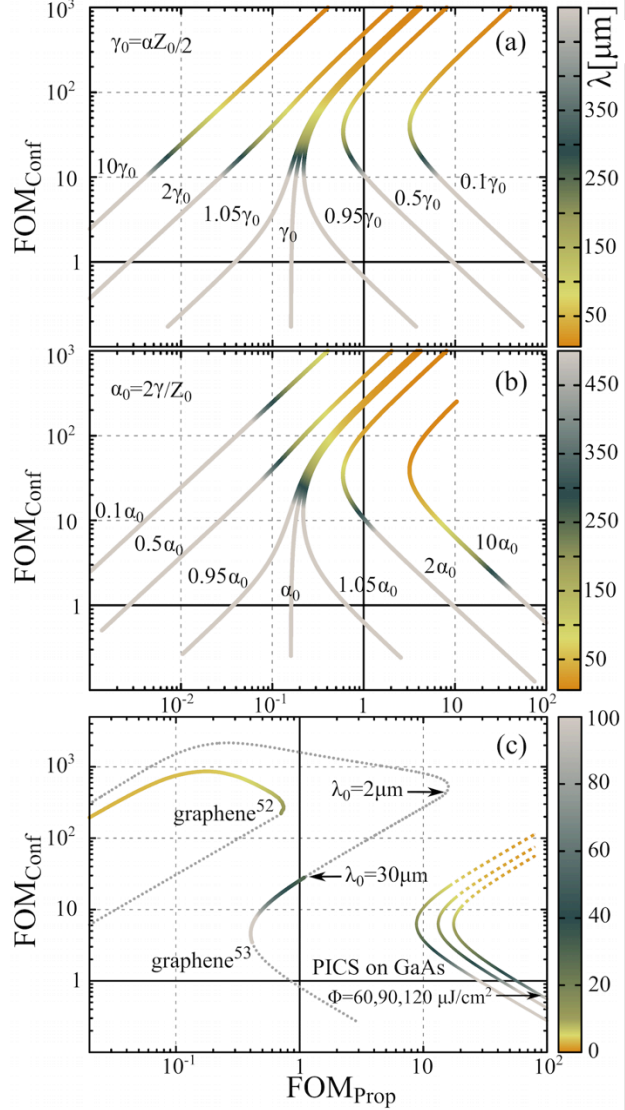

**Figure 4 | FOM relations for two-dimensional conducting systems. a**, Dependence on the collision frequency γ. **b**, Dependence on the carrier concentration α. **c**, FOMs for two experimental data series of graphene[52,53], and a photon-induced conductive sheet (PICS) on GaAs. Dotted gray lines are extrapolated data for graphene, and dashed part of the GaAs FOM shows the region close to the sheet model breakdown. For the sake of comparison the sheets are assumed to be free-standing. Color bar units are all in μm.

The current sheet approximation is valid as long as the effective thickness of the conductive sheet is negligible compared with the free-space wavelength of the incident plane-waves and compared to the skin depth of the material.

Assuming for a moment a free-standing sheet with a Drude response, $\sigma(\omega) = \alpha/(\gamma - i\omega)$, we can obtain a very simple expression for the confinement FOM,

$$\text{FOM}_{\text{conf}} = \frac{1}{\pi Z_0 \alpha}\omega \qquad (4)$$

$Z_0$ is the free-space impedance. The confinement depends linearly on the frequency of operation and it is independent of the collision frequency γ (figure 4a). On the other hand, the inverse proportionality of the confinement to α means that higher carrier concentration/stronger doping will reduce the confinement, while boosting the propagation length (see figure 4b). At $\alpha Z_0 = 2\gamma$, the low-frequency tail of the curve switches its behaviour from ascending to descending $FOM_{Prop}$ (see reference 51 for more details). However, the high-frequency tail, starting from approximately $\omega = \alpha Z_0$, shows a growing trend for both FOMs. The limiting point for this behaviour is determined by the breakdown of the sheet model and/or the onset of the interband transitions, which bend the curve towards lower propagation FOMs. Figure 4c compares figure-of-merits calculated from the measured and extrapolated data for several cases of free-standing conductive sheets[i]. Two different experimental reports[52,53] are used to generate graphene data. Noteworthy, for the former case a Drude fit is used to extrapolate the conductivity of graphene to lower frequencies, while for the latter, the effect of the interband transition calculated from Kubo formula, assuming a Fermi energy of $E_F = 600$ meV at room temperature (T = 300K)[51], is superimposed on the provided Drude fit. We see that both samples accommodate very strongly confined plasmon modes, but with limited propagation length. From the extrapolation of Yan et al. graphene data[53] to higher frequencies it seems higher propagation lengths might be possible, if the Fermi level could be increased that much without affecting the collision frequency. This is a promising challenge for experimental investigation. A more realistic case would include a substrate (see reference [51] for details and examples).

A photon-induced conductor, such as optically pumped GaAs, can also support surface plasmons. In such systems, the carrier concentration is easily modified by varying the pump fluence, creating a possibility for tunable excitation of surface plasmons[44,54]. Thickness of the induced conductive layer over which the conductivity can be considered uniform is approximately 1 μm. FOM graphs of optically pumped GaAs sheets for a few different pump fluences are shown in figure 4c. We find remarkably good figure-of-merit in the linear region for small wavelength, i.e., $FOM_{prop} = 20.058$ and $FOM_{conf} = 3.718$ at $\lambda_0 = 15$ μm for the maximum pump fluence of $\Phi = 120$ μJ/cm$^2$. For even smaller wavelengths, a two-dimensional current sheet can no longer accurately model the 1-μm-thick current layer, as we will see in the next section.

---

[i] The effect of a dielectric substrate is discussed in the supplementary information[51]. It is important to keep conditions the same in order to preserve comparability between different materials.

**THIN FILMS**

Once the thickness of the conductive sheet is increased, the sheet model discussed above breaks down and the use of a multi-layer model becomes necessary[51]. For an optically thin film, SPP modes appearing at the film boundaries hybridize into two SPP modes exhibiting a symmetric or antisymmetric nature with respect to the oscillating current direction at the interfaces (IMI, see figure 1c). For large film thicknesses, the symmetric mode (figure 5a) converges to the single-interface SPP, while reducing the thickness pushes the dispersion curve of this mode away from the light line to become more subwavelength and resemble a conductive sheet SPP. Although this improves the overall confinement of the mode, it is not a very practical case because the low-frequency tail lies at sub-unity propagation FOMs, and the high-frequency part, which has better confinement FOMs, is prone to interband transition losses. The antisymmetric mode, on the other hand, shows considerably better overall propagation FOM, yet with inferior confinement for thinner films. This is expected since the dispersion curve[51] closely follows the light-cone.

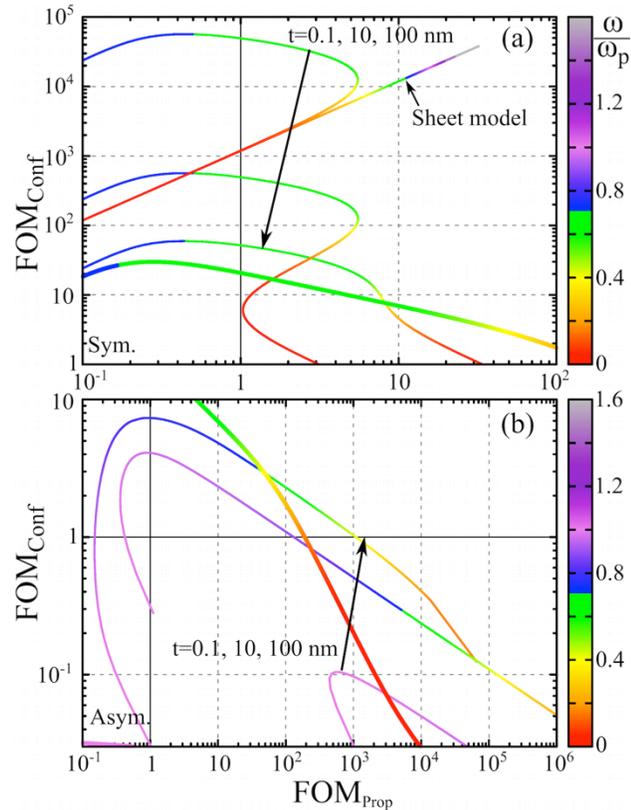

**Figure 5 | The FOM graphs for thin metal films. a**, symmetric **b**, asymmetric modes of the IMI structure for a conductor with Drude response. The thick line corresponds to the single-interface SPP solution of the same Drude model. Calculations for a sheet model using a sheet conductivity of a fictitious 0.1nm-thick conductive film is superimposed on **a**. The color bar shows the normalized frequency $\omega/\omega_p$.

It should be noted that selected thicknesses in Figure 5 are just for the sole demonstration purpose and do not necessarily present an experimental condition. Nevertheless, it provides insight into the overall behaviour of the thin-film SPPs. Decreasing the thickness of a realistic film down to few nano-meters would have a multi-fold impact on the collision frequency, modifying the Drude response to larger loss in the process. Taking this fact into account, the plotted curves present an underestimated loss for the smaller film thicknesses, which would have shifted the curves to left, further deteriorating $FOM_{prop}$. Moreover, fabrication of very thin metal films is experimentally challenging, since the films tend to break up in discontinuous islands beyond a certain thickness[55]. At 0.1nm, the slab would represent a single atomic layer. Regardless of practicality, it would be insightful to compare such a case to a sheet model of the same conductivity. A substrate will affect confinement and propagation length. Again, it is important to compare different material films under same conditions (thickness, substrate). Influence of substrate is discussed in reference 51.

**METAL-INSULATOR-METAL GEOMETRIES**

Another important geometry of choice for practical applications is a dielectric slab between two optically thick conductors, known as a metal-insulator-metal (MIM) structure. The importance of the MIM structure is highlighted by its ability to confine light in an extreme subwavelength gap region, allowing for interesting applications, such as ultracompact lasers, and chip-scale optical devices[56–58]. Similar to the IMI structure, the surface-plasmon polariton modes of the MIM structure exhibit symmetric and antisymmetric nature with their maximum energy density inside the dielectric gap region. Such modes are often referred to as gap modes. The thickness-dependent performance of the gap plasmons for a Drude metal and different gap thicknesses (100, 500 and 1000nm) is depicted in the FOM graphs of figure 6. The modes are separated with respect to their field symmetry. Compared to the single-interface waveguide (shown by the thick line in figure 6), significant enhancement of the confinement figure-of-merit can be observed for the antisymmetric mode, especially at the low-frequency part of the spectrum. However, the trade-off between confinement and propagation-length is notable in the FOM graphs. The symmetric SPP mode of the MIM waveguide converges to the antisymmetric mode near the surface plasmon frequency $\omega_{spp}$, however, it rapidly approaches and crosses the light line to reach a cut-off at lower frequencies (see S1 and S3 on figure 6). Poorer confinement is the general characteristic of the symmetric mode, particularly at lower frequency tail, which can be associated to its leaky nature inside the light-cone. For a thin gap, the symmetric mode lies between $\omega_{spp}$ and $\omega_p$, and becomes extremely lossy.

In addition to SPP modes, MIM geometry supports waveguide modes, exhibiting a cut-off frequency, which is inversely proportional to the width of the dielectric gap. By widening the gap, different propagating orders of the waveguide modes appear in the selected frequency range above the light-cone. For such modes, the field maximum is located inside the waveguide rather than at the interfaces. Despite their longer propagation length compared to the SPP modes, they suffer from mediocre confinement and need thicker dielectric gaps to show up in the frequencies close to $\omega_{spp}$. It must be noted that the geometry dependence discussed above applies the same way to all of the (bulk) materials, hence it constitutes an independence of parameter for optimization. It is also true that different geometry can compensate in performance for different material choices. The merits of different materials, thus, should be evaluated in the same geometry to be meaningfully comparable—the simplest such case is the single interface.

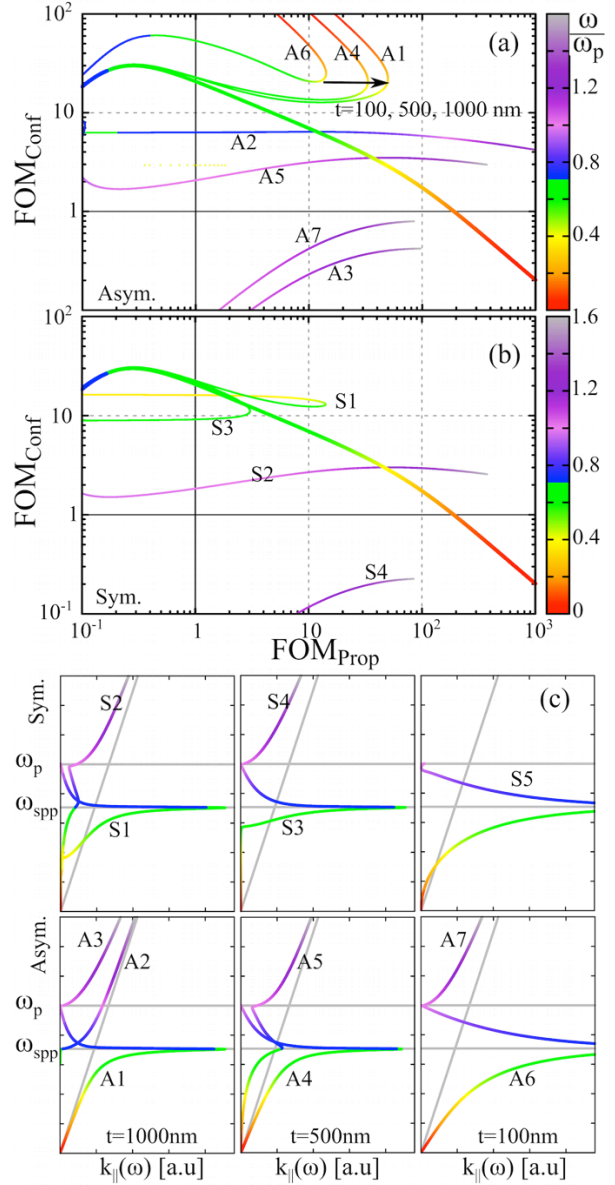

**Figure 6 | Thickness dependence of the figure of merit relations for MIM/gap modes of a Drude type dielectric function.** Modes are calculated for three different thicknesses of the gap, 1000nm, 500nm and 100nm, and include the SPP and non-SPP guided solutions for **a**, asymmetric current distribution. **b**, symmetric current distribution. **c**, Dispersion relations separated based on the mode symmetry and the gap thickness. The color bar shows the normalized frequency $\omega/\omega_p$.

## DISCUSSION

In this Article, we have argued that plasmonic materials cannot be assessed by a single number because of the inherent trade-off between the degree of confinement and the propagation length.

Rather one needs to make a two-dimensional graph of two figures-of-merits, one for the degree of confinement and one for the propagation length of surface plasmon polaritons. We have demonstrated that a two-dimensional graph of these two figures of merits allows for a clear and fair comparison of different plasmonic materials and geometries. Specifically, we find that silver is having the best properties overall, i.e., for a given degree of confinement, it always has the plasmons with the longest propagation length, except in the far-infrared where silicon carbide outperforms silver and in the ultraviolet range where aluminium outperforms silver. Other materials, including gold, TCOs and metal nitrides may be useful for other reasons, such as tunability and material compatibility. When designing a specific plasmonic application, the FOM graphs also enable to easily pick the most suitable material.

**Acknowledgments**

Work at Ames Laboratory was partially supported by the U.S. Department of Energy, Office of Basic Energy Science, Division of Materials Science and Engineering (Ames Laboratory is operated for the U.S. Department of Energy by Iowa State University under contract No. DE-AC02-07CH11358), by the U.S. office of Naval Research, award No. N00014-10-1-0925 (simulations). The European Research Council under the ERC Advanced Grant No. 320081 (PHOTOMETA) supported work (theory) at FORTH.

**Additional information**

The authors declare no competing financial interests.